#Comment: LaTeX file
-------------------------------------------

\documentstyle[preprint,aps]{revtex}
\begin{document}
\draft
\sloppy
\scrollmode

\newcommand{\lfrac}[2]{#1/#2}
\renewcommand{\baselinestretch}{1.3}       %

\title{\vspace{-5.5cm} ~ \hfill {\normalsize FUB-HEP/94-5}\\[4.5cm]
Comment on Path Integral Derivation of Schr\"odinger Equation
 in Spaces with Curvature and Torsion}
\author{P.~Fiziev\thanks{Permanent address:
 Department of Theoretical Physics,
 Faculty of Physics, University of Sofia,
 Bull.~5 James Boucher, Sofia~1126,
 Bulgaria.
 Work supported by the Commission of the
 European Communities for Cooperation in Science
 and Technology, Contract~No. ERB3510PL920264
 }
 and H.~Kleinert}
\address{Institut f\"ur Theoretische Physik,\\
          Freie Universit\"at Berlin\\
          Arnimallee 14, D - 14195 Berlin}
\date{\today}
\maketitle
\begin{abstract}
We present
 a derivation of the
 Schr\"odinger equation
for
a path integral of a point particle
in a space with curvature and torsion
which is considerably shorter and more elegant
than what is commonly found in the literature.
\end{abstract}
\newpage
\section{Introduction}
When studying the historic paper
on path integrals
in spaces with curvature and torsion
by DeWitt \cite{4}, or any of ist successors \cite{5,6,7},
one may rightfully be frustrated by
the tedious algebra
involved in deriving
the simple Schr\"odinger equation
satisfied by a
point particle
in curved space.
If torsion were to be admitted  to the geometry,
this derivation would become even more involved.

In this note, we want to point out
that there exists a much shorter and more elegant
derivation
based on the use of nonholonomic coordinate transformations.
By analogy with the physics of defects,
these have recently been found to
be an essential tool in {\em predicting\/}
the correct path integral in such spaces \cite{1,5p,2}.
The use of such transformations is
very fundamental, and the transformation
rules
constitute a new {\em quantum equivalence principle\/} \cite{1,5p}
which extends Einstein's famous classical principle
to spaces with curvature and torsion,
specifying the dynamical laws of
both classical \cite{8} and quantum physics \cite{2}.

Here we do not intend to
convince the reader of the virtues of this principle
but want to direct his attention upon a useful technical advantage
of using such transformations:
a drastic simplification of
the above-mentioned derivation of the
Schr\"odinger equation.

\section{Schr\"odinger Equations from Path Integral}
Let us first recall the simple derivation
of the
Schr\"odinger equations from path integral
of a free nonrelativistic particle of mass $M$ in euclidean space.
The set of fluctuating paths in $D$
dimensions is
parametrized by the
time-dependent cartesian
coordinates
$x^i(t)$ $(i=1,\dots,D)$. The time axis
is sliced into $N+1$ intervals $(t_n,t_{n-1})$ $(n=1,\dots,N+1)$
of width $ \Delta t$
and the
positions $x^i(t_n)$
are denoted by $x_n^i$.
If $  \Delta x^i_n$ are the differences
$  x^i_n-x^i_{n-1}$,
the time-sliced path integral
has the measure
\begin{equation}
 \int d\mu _x^{\rm new} : =  \lim_{N\rightarrow  \infty}
   \left[ \prod^{N+1}_{n=2} \int \frac{d^D ( \Delta x_n)}
      {(2 \pi i \hbar  \Delta t/M)^{D/2} }\right]  \exp
   \left[ \frac{i}{\hbar}  \sum_{n=1}^{N+1}M \frac{( \Delta x^i_n)^2}
   {2 \Delta t} \right],
\label{2}\end{equation}
Separating the last integral over  $\Delta x_{N+1}$
from the others,
the amplitude $\psi_t(x)$ at the time $t$
is seen to satisfy
the recursion relation
\begin{eqnarray}
  \psi_t (x) & = & \int \frac{d^D  (\Delta x)   }
   {(2 \pi i\hbar  \Delta t/M)^{D/2} }\exp \left[-\frac{M}
   {2i\hbar  \Delta t }( \Delta x^i)^2 \right]
 \psi_{t- \Delta t}(x- \Delta x) .
\label{000}\end{eqnarray}
Expanding
$ \psi_{t- \Delta t}(x- \Delta x)$
in powers of $ \Delta x$ and using the Gaussian integral
formula
\begin{eqnarray}
 &&\int \frac{d^D  \Delta   }{(2 \pi \varepsilon/a)^D}
 \exp
     \left[ -\frac{a}{2\varepsilon}  \left( \Delta ^ i  \Delta ^ i
\right)\right]
      \left(1 + b_i  \Delta ^ i  + b_{ ij }
      \Delta ^ i   \Delta ^ j  + \dots \right)
=    1 + \frac{\varepsilon}{a} b_{ ii }
        + {\cal O} (\varepsilon^2 ),
\label{60}\end{eqnarray}
one finds
\begin{eqnarray}
  \psi_t (x) & = & \int \frac{d^D (\Delta x) }
   {(2 \pi i\hbar  \Delta t/M)^{D/2} }
\exp \left[i\frac{M}
   {2\hbar  \Delta t }( \Delta x^i)^2 \right] \nonumber \\
&&~~~~~\times\left[ \psi_{t- \Delta t} (x) -
    ( \Delta x)^i \nabla_i  \psi_{t- \Delta t} (x)
 + \frac{1}{2} ( \Delta x)^i   (\Delta x)^j \nabla_i\nabla_j \psi_t (x) + {\cal
O}(( \Delta t)^{3/2})\right] \nonumber \\
  & = &  \psi_t (x) +  \Delta t \left[ -\partial _t
      \psi_t (x) + \frac{i\hbar }{2M}
 {\mbox{\boldmath $\nabla$}}^2 \psi_t (x)
\right] + {\cal O} ( (\Delta t)^{2}).
\label{0001}\end{eqnarray}
In the limit $ \Delta t \rightarrow 0$, this yields
the Schr\"odinger equation
\begin{equation}
 i\hbar \partial _t \psi = -\frac{\hbar ^2}{2M}  \Delta \psi.
\label{50}\end{equation}
where $  \Delta \equiv {\mbox{\boldmath $\nabla$}}^2$ is the Laplace operator.

In noneuclidean spaces, the derivation becomes complicated.
For the sake of generality, we shall admit some kind of curvature and torsion.
Let $q^\mu(t)~(\mu=1,\dots D)$ be the paths in such a general
space
${\cal S}_q $.
A
nonholonomic
transformation
\begin{equation}
  \dot x^i = e^i_\mu  (q) \dot q^\mu
\label{1}\end{equation}
maps them into a reference space ${\cal S} _x $
of ${x}^i$-vectors
$(i=1,\dots,D)$
 with
a euclidean metric.
Under this transformation, the
measure of
path integration (\ref{2}) goes over into
\cite{2}
\begin{equation}
 \int d\mu _q =  \lim_{N\rightarrow  \infty}
   \left[ \prod^{N+1}_{n=2} \int \frac{d^D ( \Delta q_n)}
      {(2 \pi i \hbar  \Delta t/m)^{D/2} }
       \frac{\partial ( \Delta x_n)}{\partial ( \Delta q_n)}   \right]  \exp
   \left( \frac{i}{\hbar}   \Delta {\cal A}_q^{N+1}\right),
\label{3}\end{equation}
where
$ \Delta {\cal A}_q^{N+1}$ is the time-sliced version of the
classical action
\begin{equation}
  {\cal A}_q = \int^{t_2}_{t_1}   dt
\frac{m}{2} g_{\mu \nu} (q)
	\dot q^ \mu \dot q^ \nu\label{4}\end{equation}
 of the system, evaluated along the classical orbits.
The tensor $g_{\mu \nu } (q) = \sum_i e^i_ \mu (q)
  e^i_  \nu (q)$ is
 the  metric with a nonzero Riemann
curvature tensor
$\bar R_{ \mu \nu \lambda }{}^ \kappa $ derived from the covariant curl of the
usual
Christoffel symbols (Levi-Civita connection).
The action  $\Delta {\cal A}_q^{N+1}$
can be expanded around prepoint, midpoint, or postpoint
in each time slice,
with the latter
being the most convenient one
for the derivation of the
Schr\"odinger equation \cite{1,2}.
It will be  denoted by  $ \Delta {\cal A}_>^{N+1}$.

By a straight-forward but tedious generalization of
the above euclidean derivation
one finds
 in the space ${\cal S}_q$
 the Schr\"odinger equation
\cite{4,5,6,7,1,2}
\begin{equation}
 i\hbar \partial _t \psi = -\frac{\hbar ^2}{2m}  \Delta \psi
\label{5}\end{equation}
where $ \Delta\equiv  D^ \mu D_ \mu $ is the Laplace operator
in a general metric-affine space. This operator is
related to the
Laplace-Beltrami operator $\bar \Delta \equiv
\bar D^ \mu \bar D_ \mu\equiv  \sqrt{g}^{-1}\partial _\mu \sqrt{g}
g^{\mu \nu}
\partial _ \nu$
by
$D^ \mu D_ \mu = \bar \Delta  - 2S^\mu\partial _\mu $.
The symbol $D_ \mu $ denotes the covariant derivative with respect
 to the affine connection $ \Gamma _{ \mu  \nu  }{}^ \gamma
 = e_i{}^ \gamma  \partial _ \mu  e^i_ \nu $ with nonzero
torsion $S _{ \mu  \nu  }{}^ \gamma  =  \Gamma _{[ \mu
\nu ]}{}^ \gamma $, and zero Cartan curvature
$R_{ \mu  \nu  \lambda }{}^  \kappa = 0$,
whereas
$\bar D_ \alpha $ is the covariant derivative with respect to the
Christoffel symbols $\bar  \Gamma _{ \mu  \nu }{}^ \kappa =  g^ { \kappa
\lambda }\left(\partial _
\mu g_{ \nu  \lambda }
 + \partial _ \nu g_{ \mu  \lambda } - \partial _
\lambda g_{ \mu  \nu }
\right)/2$ with nonzero Riemann curvature $\bar R_{ \mu  \nu  \lambda }{}
^  \kappa $,
and zero
torsion $\bar S _{ \mu  \nu  }{}^ \gamma  = \bar \Gamma _{[ \mu
\nu ]}{}^ \gamma = 0$.

What makes the historic
 derivation \cite{4,5,6,7} of the Schr\"odinger equation~(\ref{5})
tedious is the fact,
observed
first
in Ref. \cite{3}, that one has
to calculate the time-sliced
 action $ \Delta {\cal A}_>^{N+1}$
up to fourth-order terms
in the differences $ \Delta q_n^ \mu  = q_n^ \mu  -
q_{n-1}^ \mu $,
and the Jacobian $\lfrac{\partial ( \Delta x_n)}
 {\partial ( \Delta q_n)}$ up to the second-order.
Since $ \Delta q_n^\mu$ are of the order $ \sqrt{ \Delta t}$,
all these terms contribute
to first order in $ \Delta t$,
and thus to the Schr\"odinger equation.
Instead of
(\ref{60}), the derivation requires the more general
formula
\begin{eqnarray}
 &&\int \frac{d^D  \Delta   }{(2 \pi \varepsilon/a)^D}\sqrt{g}
      \left(1 + b_ \mu  \Delta ^ \mu  + b_{ \mu  \nu }
      \Delta ^ \mu   \Delta ^ \nu  + \dots \right)\nonumber \\
&&~~~~~~~~~\times \exp
     \left[ -\frac{a}{2\varepsilon}  \left(g_{ \mu  \nu }
      \Delta ^ \mu  \Delta ^ \nu +G_{ \mu  \nu  \lambda }
   \Delta ^ \mu   \Delta ^ \nu  \Delta ^  \lambda + G_{ \mu  \nu
      \lambda  \kappa }  \Delta ^ \mu  \Delta ^ \nu  \Delta ^ \lambda
     \Delta ^ \kappa  + \dots \right)\right] \nonumber \\
    & & ~~~~~~~~~ =    1 + \frac{\varepsilon}{a} \left[ b_{ \mu  \nu }
	     g^{ \mu  \nu }  -\frac{1}{2} \left(G_{ \mu  \nu
		  \lambda  \kappa } + G_{ \mu  \nu  \lambda }
       b_ \kappa \right) g^{ \mu  \nu  \lambda  \kappa  }+
       \frac{1}{8} G_{\alpha  \beta  \gamma }G_{\mu  \nu  \lambda }
       g^{ \alpha  \beta  \gamma  \mu  \nu  \lambda }\right]
       + {\cal O} (\varepsilon^2 ).
\label{6}\end{eqnarray}
  Here
 $b_ \mu , b_{ \mu  \nu }$,  and $g_{ \mu  \nu },
   G_{ \mu  \nu  \lambda }, G_{ \mu  \nu  \lambda  \kappa },
 \dots $ are $\varepsilon$-independent coefficients,
  $g $ denotes the determinant of the metric $\det \| g_{ \mu  \nu }\| $,
and $g^{\mu_1,\dots,\mu_{2n}}$ are symmetric tensors formed
from products of $n$ metric tensors:
 $g^{ \mu  \nu  \lambda  \kappa } = g^{ \mu  \nu }
g^{ \lambda  \kappa } + g^{ \mu  \lambda } g^{ \nu  \kappa }
 +  g^{ \mu  \kappa } g^{ \nu  \lambda },~~ g^{ \alpha  \beta
      \gamma \mu  \nu  \lambda } = g^{ \alpha  \beta }
   g^{ \gamma \mu  \nu  \lambda } + g^{ \alpha  \gamma }
   g^{ \beta \mu  \nu  \lambda }+ \dots ~$.
In the course of the calculations
one encounters
many  cancelations, which
make the final result  (\ref{5})
again very simple.
There must be
a derivation which  reflects the simplicity
of the final result from the beginning, and this is what
we want to present now.

\section{New Derivation}
First we observe that
the time-sliced action
$ \Delta {\cal A}^{N+1}_q $ which consists
of the
$\sum_{n=2}^{N+1}  \Delta  {\cal A}_n $
calculated along the classical trajectories
is a simple Gaussian when expressed
in terms of the velocities $\dot q^\mu$ in each time slice.
The classical trajectories
are described by the equation of motion
\begin{equation}
   \ddot q^ \mu  +  \Gamma _{ \nu \lambda }{}^ \mu  \dot q^ \nu\dot q^  \lambda
= 0.
\label{7}\end{equation}
This equations implies the conservation
of the energy
 $E =  g_{ \mu  \nu } \dot q^ \mu
    \dot q^ \nu /2 $.  As a consequence, the short-time
action is
\begin{equation}
   \Delta {\cal A} = \frac{m}{2} g_{ \mu  \nu } \dot q^{ \mu }
  \dot q^ \nu   \Delta t = \frac{m}{2 \Delta t} g_{ \mu  \nu }
    \Delta\xi^ \mu  \Delta\xi^ \nu ,
\label{8}\end{equation}
where we have found it useful to introduce the {\em vector\/} quantities
$ \Delta\xi^ \mu$ defined by
$ \Delta\xi^ \mu  \equiv \dot q ^ \mu
 \Delta t$ whose size is of the order of $( \Delta t)^{1/2}$, to have
quantities comparable to the
previous differences $ \Delta x^i$.
In (\ref{8}),
the velocities $\dot q^\mu$ as well as the metric
 $g_{ \mu  \nu }$ are calculated at
the latest time $t_n$ in the interval $(t_n,t_{n-1})$,
to have the preferred post-point form (any time would give the same result, due
to energy conservation).

An explicit {\em functional} relation between
$\Delta \xi^\mu$ and $ \Delta q^\mu$ is obtained by Taylor-expanding
\begin{eqnarray}
\Delta q^ \mu & =& q^ \mu (t_n)
  -q^ \mu (t_{n-1}) =  q ^\mu (t_n) - q^ \mu  (t_n- \Delta t)
= \dot q ^ \mu  \Delta t - \frac{( \Delta t)^2}{2!}\ddot q^ \mu +
 \frac{( \Delta t)^3}{3!}\dot{\ddot {q}}^ \mu + {\cal O} ( (\Delta t)^4)
\nonumber \\
&&
\label{}\end{eqnarray}
 and using (\ref{7}),
which implies for the higher time derivatives
$$\ddot q  ^\mu= -  \Gamma _{\nu \lambda } {}^ \mu\dot q^ \nu \dot q ^  \lambda
,~~~~
 \dot{\ddot q}^ \mu = -\left(\partial _  \kappa
 \Gamma _{ \nu  \lambda}{}^ \mu
 -2  \Gamma _{\{  \kappa  \delta\} }{}^ \mu
\Gamma _{ \nu \lambda }{}^  \delta
\right)\dot {q}^  \kappa  \dot {q}^ \nu \dot {q}^ \lambda,   $$
where curly braces around indices indicate their symmetrization.
Hence,
%
\begin{eqnarray}
  \Delta q^ \mu & = &\Delta \xi^ \mu  + \frac{1}{2!} \Gamma _{\nu  \lambda }{}
^ \mu  \Delta \xi^\nu  \Delta \xi^ \lambda  - \frac{1}{3!} \left(\partial _
\kappa
   \Gamma _{\nu  \lambda } {}^ \mu  - 2  \Gamma _{\{  \kappa  \delta \}}{}
  ^\mu   \Gamma _{\nu  \lambda }{}^ \delta \right)\Delta \xi ^ \kappa
     \Delta \xi^\nu  \Delta \xi^ \lambda  +  \dots ,
   \label{9.a}
\end{eqnarray}
and this may be inverted to
\begin{eqnarray}
 \Delta \xi^ \mu & = &  \Delta q^ \mu  - \frac{1}{2!}  \Gamma _{\nu  \lambda
}{}^ \mu
   \Delta q^\nu   \Delta q^ \lambda  + \frac{1}{3!} \left(\partial _ \kappa
     \Gamma _{\nu  \lambda }{}^ \mu  +  \Gamma _{\{  \kappa   \delta \}}
   {}^ \mu     \Gamma _{\nu  \lambda }{}^  \delta \right)
    \Delta q^ \kappa   \Delta q^\nu   \Delta q^ \lambda  +  \dots \,\,\, .
\label{9.b}\end{eqnarray}
 Using this equation, we change
 the integration variables $ \Delta q_n^ \mu$ in formula
 (\ref{3})  into $\Delta \xi_n^\mu$,
and find the following measure for the path integral
\begin{equation}
  \int d\mu _q = \lim_{N \rightarrow   \infty} \left[
 \prod_n \frac{d^D\Delta \xi_n}{
(2 \pi i \hbar  \Delta t_n/m)^{D/2}}
\sqrt{g(q_n)}\right]
 \exp \left[ \frac{i}{\hbar  \Delta t}\sum_{n=1}^{N+1}
   g_{ \mu  \nu } (q_n) \Delta \xi_n ^ \mu  \Delta \xi_n^ \nu \right] .
\label{10}\end{equation}
 This expression is related to
the flat-space measure
(\ref{2})
 by just a linear transformation.
The reason for this is that in terms of the nonholonomic $x$-variables,
the  equation of motion (\ref{7})
is
trivial: $\ddot x^i = 0$. Hence $\dot {x}^ i = {\rm const}$, and
\begin{equation}
 \Delta x^i = \dot x^{i}  \Delta t =  e^i_\mu  \dot q^\mu   \Delta t =
 e^i_\mu  \Delta \xi^\mu      ,
 \label{}\end{equation}
so that
 $ \lfrac{\partial ( \Delta x)}{\partial (\Delta \xi)  }  = \det \| e^i_\mu \|
=  \sqrt{g} $.

Using the simple measure (\ref{10}),
we immediately find
for an amplitude $\psi_t(q)$
the recursion relation
 (again by removing the last slice from the product of
integrals)

\begin{eqnarray}
  \psi_t (q) & = & \int \frac{d^D \Delta \xi }
   {(2 \pi i\hbar  \Delta t/M)^{D/2} } \sqrt{g} \exp \left(-\frac{M}
   {2i\hbar  \Delta t }g_{ \mu  \nu } \Delta \xi^ \mu
   \Delta \xi^ \nu \right)
 \psi_{t- \Delta t}\left(q- \Delta q(\Delta \xi)\right) .
\end{eqnarray}
Expanding the amplitude inside the integral
in powers of $ \Delta q^\mu$, and expressing these in terms of
 $\Delta\xi^\mu $ with the help of (\ref{9.a}),
we find
\begin{eqnarray}
  \psi_t (q) & = & \int \frac{d^D \Delta \xi  }
   {(2 \pi i\hbar  \Delta t/M)^{D/2} } \sqrt{g}\exp \left(-\frac{M}
   {2i\hbar  \Delta t }g_{ \mu  \nu } \Delta \xi^ \mu
   \Delta \xi^ \nu \right) \nonumber \\
&&\!\!\!\!\!\!\!\!\!\!\!\!\!\!\!
\times\left[ \psi_{t- \Delta t} (q) -
    \left(\Delta \xi^\mu  + \frac{1}{2}  \Gamma _{\nu \lambda   }
  {}^\mu  \Delta \xi^\nu  \Delta \xi^\lambda  \right)\partial _\mu  \psi_{t-
\Delta t} (q)
 + \frac{1}{2} \Delta \xi^\mu  \Delta \xi^ \nu \partial _{\mu}\partial _  \nu
  \psi_t (q) + {\cal O}(( \Delta t)^{3/2})\right] \nonumber \\
  & = &  \psi_t (q) +  \Delta t \left\{ -\partial _t
      \psi_t (q) + \frac{i\hbar }{2m} \left[g^{\mu  \nu }
   \partial _{\mu }\partial _{ \nu } \psi_t (q) - \Gamma _ \kappa{}
^{ \kappa \mu }
   \partial _\mu \psi_t (q)\right]\right\} + {\cal O} ( (\Delta t)^{2}).
\label{gaussi}\end{eqnarray}
After making use of the identity $g^{\mu  \nu } \partial _{\mu}
\partial _{  \nu }-  \Gamma _ {\kappa }{}^{ \kappa \mu }\partial _\mu =
 D^\mu  D_\mu    $, this is precisely
the Schr\"odinger equation (\ref{5}).
\section{Comparison with the historic derivation of the Schr\"odinger equation
 in spaces with curvature and torsion }
 To compare the present derivation
with the historic
derivation in Refs. \cite{4,5,6,7,1,2}, we
 must calculate $ \Delta  {\cal A} $ as a function of
$ \Delta q^ \mu  $ up to fourth order terms
using formulas
(\ref{8}) and (\ref{9.b}). This yields
\begin{eqnarray}
  \Delta  {\cal A} & = & \frac{m}{2 \Delta t}\left[ g_{\mu  \nu }
                   \Delta q^\mu   \Delta q^ \nu  - \bar
		  \Gamma _{\mu  \nu  \lambda }  \Delta q^\mu
        \Delta q^ \nu   \Delta  q^ \lambda  + \frac{1}{3}
        \left(\partial _\mu  \bar  \Gamma _{ \nu  \lambda  \kappa }-
     \frac{1}{4} \bar \Gamma _{\mu  \nu  \sigma } \bar
    \Gamma _{ \lambda  \kappa }
 {} ^ \sigma \right)  \Delta q ^\mu   \Delta q^ \nu   \Delta q^ \lambda
    \Delta q^ \kappa \right. \nonumber \\
   &&~~~~~~~~~\left. + \frac{1}{3} S^ {\sigma }{}_{\mu  \nu }
         S_ {\sigma  \lambda  \kappa }
      \Delta q^\mu   \Delta q^ \nu  \Delta q^ \lambda
        \Delta q^ \kappa  + {\cal O} (( \Delta q)^5)\right] .
\label{11}\end{eqnarray}
The Jacobian
$ \lfrac{\partial ( \Delta x)}{\partial  ( \Delta q)}$
is conveniently given in the exponential form
\begin{equation}
\lfrac{\partial ( \Delta x)}{\partial  ( \Delta q)} =  \sqrt{g}  \exp
\left({i}  \Delta {\cal A}_J'/\hbar\right),
\label{expf}\end{equation}
with an effective action
\begin{equation}
  \frac{i}{\hbar }  \Delta {\cal A}_J' = -  \Gamma _{\{\mu  \nu  \}}{}^\mu
        \Delta q^ \nu  + \frac{1}{2} \left[ \partial _{\{\mu }
    \Gamma _{ \lambda  \nu \} }{}^\mu + \frac{1}{3}
    \left( \Gamma _{\{ \mu  \kappa \}} {}^ \mu
    \Gamma _{\{ \lambda  \nu \}} {}^ \kappa -
  \Gamma _{\{  \lambda  \kappa \} }{}^\mu   \Gamma _{\{  \nu \mu \}}{}^ \kappa
   \right)\right]   \Delta q^ \lambda   \Delta q ^ \nu+{\cal O}
  (( \Delta q)^3).
\label{12}\end{equation}
Inserting these expansions into the product of integrals
(\ref{3}) we obtain a large number of terms.
Upon applying
formula (\ref{6}),
most of these cancel each other,
leading again to the
 Schr\"odinger
equation (\ref{5}).

The additional labor is the same as if we were to prove
the obvious
identity
\begin{equation}
[\lfrac{\partial ( \Delta x)}{\partial ( \Delta q)}]    [\lfrac{\partial (
\Delta q)}
 {\partial (\Delta \xi) } ] = \lfrac{\partial  ( \Delta x)}{\partial (\Delta
\xi)} =  \sqrt{g}
\label{iden}\end{equation}
by writing
$
[\lfrac{\partial ( \Delta x)}{\partial ( \Delta q)}]
$
as in (\ref{expf})
expanded via (\ref{12}), and
writing
 $ \lfrac{\partial ( \Delta q)}{\partial (\Delta \xi)} = \exp \left({i}
\Delta  {\cal A}_J''/\hbar\right) $
with
\begin{equation}
 \frac{i}{\hbar }  \Delta  {\cal A}_J'' =  \Gamma _{\{ \mu  \nu \}}{} ^\mu
  \Delta \xi^ \nu  - \frac{1}{2} \left(\partial _{\{ \mu }
    \Gamma _{ \lambda  \nu \}}{}^\mu - 2  \Gamma _{\{\{ \mu  \kappa \}}{}^
\mu
     \Gamma _{ \lambda  \nu \}}{}^ \kappa  +  \Gamma _{\{ \mu  \lambda \}}{}
  ^  \kappa \Gamma _{\{  \kappa  \nu \}}{}^\mu \right)
   \Delta \xi^ \lambda \Delta \xi^ \nu  + {\cal O} (\Delta \xi^3),
\label{13}\end{equation}
which follows from
(\ref{9.a}).
Inserting here
(\ref{9.b})
one may reexpress the right-hand side as a power series in $ \Delta q$,
and form the
sum $  \Delta {\cal A}_J' + \Delta {\cal A}_J''$.
The many terms in this sum
all cancel each other,
as required by (\ref{iden}).

All such complications are avoided
by working with the variables $ \Delta\xi^\mu$ in which the exponential
in (\ref{gaussi}) is a pure Gaussian.



\begin{thebibliography}{11}
\bibitem{4} B.\ S.\ DeWitt, Rev.\ Mod.\ Phys.\ {\bf 29}, 377 (1957)
\bibitem{5} K.\ S.\ Cheng, J.\ Math.\ Phys.\ {\bf 13},
    1723 (1972)
\bibitem{6} H.\ Dekker, Physica, {\bf A103}, 586 (1980)
\bibitem{7} G.\ M.\ Gavazzi, Nuovo Cimento, {\bf A101}, 241 (1989)
\bibitem{1} H.~Kleinert, Mod.\ Phys.\ Lett.\ {\bf A4},
2329 (1989)
\bibitem{5p} H.~Kleinert, Phys.\ Lett.\ B {\bf 236},
 315 (1990)
\bibitem{2} H.~Kleinert, {\em Path Integrals in Quantum
 Mechanics, Statistics, and Polymer Physics\/},
  World Scientific Publ.\ Co.\ (1995)
\bibitem{8} P.\ Fiziev and H.\ Kleinert, {\em Variational
    Principle for Classical Particle Trajectories in Spaces
   with Torsion\/}, hep-th/9503074, and {\em Euler Equations for Rigid Body ---
A Case for Autoparallel Trajectories in Spaces with Torsion\/},
hep-th/9503075.
\\
H. Kleinert und A. Pelster, {\em Lagrange Mechanics in Spaces with Curvature
and Torsion\/},
FU-Berlin preprint 1996;\\
The three  papers can be read
directly on the www, the first under \\
http://www.physik.fu-berlin.de/~kleinert/kleiner\_re219/newvar.html,
the second with 219/newvar.html replaced by 224/euler.html,
the third by 243/varpr.html.
\bibitem{3} S.\ F.\ Edwards, Y.~V.\ Gulyaev, Proc.\ Roy.\ Soc.\
   London, {\bf A279}, 229 (1964)


%

\end{thebibliography}
\end{document}